\documentstyle[12pt]{article}
\input epsf

\newlength{\extraspace}
\newlength{\extraspaces}
\newcommand{\be}{\begin{equation}
\addtolength{\abovedisplayskip}{\extraspaces}
\addtolength{\belowdisplayskip}{\extraspaces}
\addtolength{\abovedisplayshortskip}{\extraspace}
\addtolength{\belowdisplayshortskip}{\extraspace}}
\newcommand{\ee}{\end{equation}}

\newcommand{\bq}{\begin{eqnarray}
\addtolength{\abovedisplayskip}{\extraspaces}
\addtolength{\belowdisplayskip}{\extraspaces}
\addtolength{\abovedisplayshortskip}{\extraspace}
\addtolength{\belowdisplayshortskip}{\extraspace}}
\newcommand{\eq}{\end{eqnarray}}

\newcommand{\newsection}[1]{
\vspace{15mm}
\pagebreak[3]
\addtocounter{section}{1}
\setcounter{equation}{0}
\setcounter{subsection}{0}
\setcounter{footnote}{0}
\begin{flushleft}
{\large\bf \thesection. #1}
\end{flushleft}
\nopagebreak
\medskip
\nopagebreak}

\begin{document}
\hbox{}
\nopagebreak
\vspace{-3cm}
\addtolength{\baselineskip}{.8mm}
\baselineskip=24pt
\begin{flushright}
{\sc OUTP}- 96  73 P\\
hep-th@xxx/9703128 \\
\today
\end{flushright}

\begin{center}
{\Large  Compact QED$_3$ with $\theta$ term and axionic confining strings.}\\
\vspace{0.1in}

{\large William E. Brown and Ian I. Kogan}
\footnote{ On  leave of absence
from ITEP,
 B.Cheremyshkinskaya 25,  Moscow, 117259, Russia}\\
{\it  Theoretical
 Physics,
1 Keble Road, Oxford, OX1 3NP, UK}\\
\vspace{0.1in}
\vspace{0.1in}

{\sc  Abstract}

\end{center}
We discuss three dimensional compact QED with a $\theta$ term due
to an axionic field.  The variational gauge invariant functional is
considered and it is shown that the ground state energy is independent
of $\theta$ in a leading approximation.  The mass gap of the axionic
field is found to be dependent upon $\theta$, the mass gap of the
photon field and the scalar potential.  The vacuum expectation of the
Wilson loop is shown to be independent of $\theta$ in a leading
approximation, to obey the area law and to lead to confinement.  We also
briefly discuss the properties of axionic confining strings.
\noindent

\vfill

\newpage
\newsection{Introduction.}

Within QFT one can add to any Lagrangian of a field with non-trivial
topology a gauge and Lorentz invariant `$\theta$ term'.  Such terms take the form of the topological charge of the theory with a coefficient of $\theta$, the
Fourier transform variable of the winding number, suitably normalized.
These terms arise naturally in any theory where there are
topologically non-trivial solutions of the classical equations of
motion i.e. instantons.  For example, see the reprint volume
\cite{Shifman} and the references therein.  The $\theta$ term
cannot be traced in a perturbation theory because it has no affect
upon the classical equations of motion.  In this paper we shall
consider the effect of the $\theta$ term in compact $2+1$ QED.

Much work has been carried out upon such theories and their associated
phenomena.  2$+$1 QED was studied by Polyakov, \cite{polyakov}, for zero $\theta$, who found a
mechanism of confinement due to non-perturbative effects caused by monopoles.  This has led
to its consideration as a simpler (Abelian) model with many of the
same characteristics as QCD.  Compact QED$_3$ with non-zero $\theta$ was
considered in a path integral approach by Vergeles \cite{verg} who
found that the long range interactions of the
monopoles and the anti-monopoles were responsible for the suppression
of the $\theta$ dependence of the vacuum energy of the theory as a
sub-leading term in volume (in the limit of large volume).  This is in
striking contrast with QCD,
where the $\theta$ dependence of the vacuum energy in the absence
of massless quarks is firmly established.  

Vergeles used the intuitive approach of a
partition function of a gas of N$+$Q monopoles and N anti-monopoles, with
screening in the bulk and all excess monopoles deposited upon the
boundary, and summed over N and Q in the limit of large volume.  He
found that the $\theta$ dependence of the free energy increased as
V$^\frac{1}{3}$ but, for large volume, that this is suppressed as
sub-leading by the $\theta$ independent part of the free energy which
increased with V.  

The work of Vergeles  was later generalized by Samuel who
proposed that, in a 3$+$1 Yang-Mills theory, if long-ranged
interactions between instantons are present the $\theta$ parameter
may relax to zero, \cite{Samuel}.  Zhitnitsky, \cite{Zhitnitsky},
also discussed these problems in 2$+$1 QED and 3$+$1
gluodynamics.  Both of these
models possess long range interactions of topological charges but
Zhitnitsky found that, because the pseudoparticles of 3$+$1
gluodynamics possess an additional quantum number (apart from the
topological charge), only in QED$_3$ does the physics not depend on $\theta$.

Recently Polyakov has used the compact U(1) theory as a model to
construct the new type of strings, the so-called confining strings,
\cite{Polyakov2}.  It is of natural interest to ask the question of
how the $\theta$ term will reveal itself in this new string theory.
In \cite{Polyakov2}, Polyakov proposed the hypothesis that all gauge
theories are equivalent to a certain non-standard string theory, where
different gauge groups are accounted for by the weights ascribed to the
world sheets of different topologies.  The string ansatz was
established in the case of the Abelian gauge groups and conjectures
were made concerning the non-Abelian generalization.  The theories
were considered in the absence of any $\theta$ terms, however, and we
find that our proposed low energy Lagrangian for QED$_3$ with a
$\theta$ term reproduces the results of \cite{Polyakov2} with a
$\theta$ dependent, but sub-leading in momentum cut-off, modification
to the photon mass and a $\theta$ dependent shift of the minimal
surface.  This modification of the photon mass is in exact
agreement with the results of our variational calculation.

We follow the gauge invariant functional variational method of \cite{var} to approach the problem of QED$_3$ with a $\theta$ term.
Initially the aim was to repeat the calculations of \cite{var} with
non-zero $\theta$.  Through the Hamiltonian formalism,
however, it became obvious that, to introduce the $\theta$ term in
QED$_3$ and to be able to solve the theory at the space-time boundary, one has to introduce an extra degree of freedom.  A scalar
field must be added whose mass is a free parameter of the theory.
There are two very natural ways to see the form that the Lagrangian
including a $\theta$ term should take.  

The first way is to see that
U(1) singular monopole solutions can be obtained from the original
Georgi-Glashow model by breaking the internal symmetry down from SU(2)
to U(1) with a triplet of scalar fields, \cite{polyakov}.  The new
U(1) theory is compact because U(1) is a subgroup of a compact SU(2)
internal symmetry group.  The masses of the charged vector bosons are a UV cut-off for this new
theory but the mass of the third scalar field is independent of this
cut-off and may even be made zero in the BPS limit.  Thus, the
low-energy U(1) theory has two degrees of freedom; the
photon and the scalar fields.  Without the $\theta$ term there is no
interaction between the scalar and the vector field.  In
the original SU(2) theory it is natural to have the scalar and vector
fields coupled within the $\theta$ term and this form is preserved
when the internal symmetry is broken to yield the desired $2+1$ U$(1)$
theory.  The term we
find is the same as that proposed by Affleck, 
Harvey and Witten for the SU(2) Georgi-Glashow model with fermions,
\cite{Affleck}.  In the limit of a very massive
scalar field, our model reduces to the purely gauge field model
studied by Polyakov and Vergeles, but it is in the BPS limit (zero
potential but non-zero vacuum expectation value for the scalar field) that the
scalar sector of our model becomes most interesting.

The second way to obtain the same Lagrangian is to consider the
compactification of a U$(1)$ theory from $3+1$ to $2+1$ dimensions.
The component of the vector field in the direction of the compactified
spatial dimension becomes the scalar field and, although the coupling
constants in the $3+1$ U$(1)$ theory are dimensionless, the source of
the mass dimensions of the coupling constants in the $2+1$ theory
becomes apparent.

Naively, one may assume that the addition of such a $\theta$ term would make
the dynamics of the fields dependent upon theta.  We have found that
the dynamically generated mass of the photon is independent of
$\theta$ in the limit of large UV
cut-off momentum in QED$_3$ with a scalar field and,
in agreement with \cite{verg}, that the vacuum energy of the photon is
independent of $\theta$ in this limit.  We have also
found that the scalar field has a dynamically generated
mass, the square of which has two terms; one is
proportional to $\theta^2$ and the other is a bare mass term involving the
coefficient of the scalar potential.  In the BPS limit the
coefficient of the scalar potential is set to zero and the scalar mass
becomes proportional to $\theta$.  The prediction
of a dynamically generated scalar mass proportional to $\theta$ is a
new result and one that could be verified through a lattice
simulation.  The $\theta$ dependence of the
scalar vacuum energy is also greatly suppressed.  In the BPS limit it
is suppressed in terms of $O(z^{\frac{3}{2}})$ where $z\sim \exp
[-\Lambda]$ and $\Lambda$ is the UV cut-off of our theory.  In the
limit of a very massive scalar field the scalar, and hence the total,
vacuum energy is independent of $\theta$ with all $\theta$ dependence
suppressed in sub-leading terms of the scalar mass.  This is in
agreement with \cite{verg}.  We
have also calculated the string tension and show that, independently
of $\theta$, this obeys the area law and so leads to confinement as in
the simple case of purely gauge field QED$_3$, \cite{var}.

In the next section we shall describe a simple, one dimensional model
which has many of the same characteristics as more complex theories,
such as QED$_3$.  This section is intended to have a pedagogical role
as the study of this simple model gives some intuition as to the
correct formalism required to write down the $\theta$ dependence of
the variational ansatz.  We are unaware of any other instances in the
literature of a term similar to the $\theta$-term we use and so we shall give some
motivation for it in the third section, where we shall also explain the
form of our Hamiltonian.  In the fourth section we give our
ansatz for the wavefunctional.  In the
fifth section we shall perform the variational calculation; examining
the $\theta$ dependence of the vacuum energies, the masses of the theory
and the string tension.  In the sixth section, the modification of the
Polyakov model of the confining string, which we shall call the
axionic string, will be considered.

\newsection{A simple one dimensional model.}

To develope some intuition for the formalism required in order to
write down the $\theta$ dependence in the variational ansatz for
the vacuum wavefunctional of more complex models, such as QED$_3$ and QCD,
we shall first observe some results from a well known and much simpler
one dimensional model.  The formalism of this section will be modified
slightly for the more sophisticated case of QED$_3$ but it  is hoped
that the reader will gain some insight from this simple example.  This
model has many of the same
characteristics as more complex theories, such as a $\theta$ term, and
is defined by
\begin{equation}
S^{\theta} = \int dt [\frac{1}{2}(\dot{\phi})^2+\theta \dot{\phi}-\lambda
\cos \phi]
\label{xx}
\end{equation}
A few results are immediately apparent from the path integral
approach.  The first is that the minima of the theory occur at $\phi =
(2n+1)\pi, n=0,\pm1,\pm2,...$.  In analogy with the Georgi-Glashow
model with zero scalar potential (but a non-zero vacuum expectation of
the scalar field which is necessary to define the non-trivial
topology of the theory) we shall continue to study
the model in the limit of zero $\lambda$.  Secondly, the second term
is a total derivative which, after explicit evaluation of the
integral, counts the solitons of the system and gives $2\pi n$ where
$n$ is the winding number of the field configuration.  Hence, this
$\theta$ term is completely analogous to those of more complex
theories and affords the symmetry $\theta \rightarrow \theta + 1$.
Thirdly, we note that (\ref{xx}) is invariant under $\phi \rightarrow
\phi + 2\pi$, which is analogous to the gauge invariance of more complex theories.

We shall proceed with the Hamiltonian formalism, having established
the similarity between this simple theory, with
its analogues of gauge invariance and a $\theta$ term with periodicity
in $\theta$, and more complex theories.  In particular, we shall see how the
periodicity of $\theta$, which is explicit in the path integral
approach, is displayed within the Hamiltonian formalism.

The Hamiltonian of the system is
\begin{equation}
H^{\theta}=\frac{1}{2}(\pi_{\tilde{\phi}}-\theta)^2
\end{equation}
where
\begin{equation}
\phi = 2\pi n +\tilde{\phi}
\label{zer}
\end{equation}
Firstly, we see that the $\theta$ term has been absorbed into a modification of
the canonical momentum of the free field, $\tilde{\phi}$, which is
connected to the field $\phi$ in momentum space by $\phi
(k) = 2\pi n \delta (k) +\tilde{\phi} (k)$.
Secondly, the Hamiltonian is independent of $n$ and so is
explicitly invariant under the transformation $n \rightarrow n + 1$.  
The invariance of the Hamiltonian under $\theta \rightarrow
\theta + 1$, however, is unclear.  We note here that the free field
$\tilde{\phi}$ fluctuates about its' vacuum expectation value of zero
which is centred at one of the minima of $\phi$.

We require the solution of the Hamiltonian equation of the
field $\phi$.  This will be obtained as a modification of the solution
of the Hamilton equation for the field $\tilde{\phi}$,
\begin{equation}
H^{\theta} \psi = E^{\theta} \psi
\label{ham}
\end{equation}
where we initially postulate the form $\psi = \psi^{\theta} \psi^0$,
with $\psi^0$ satisfying $H^0 \psi^0 = E^0 \psi^0$.  For this simple
model $\psi^0$ is known, but for more complex theories (e.g. QCD) it
is not.  We therefore have $\psi^0 = \exp [im\tilde{\phi}]$, $E^0 =
\frac{1}{2}m^2$.  When $\theta =0$, $m=0,\pm 1,\pm 2,...$ but for non-zero $\theta$ this
is no longer correct.  We will see that is the coefficient of the
field $\phi$ in the exponential of $\psi$ that must be an integer for
non-zero $\theta$.
We make the further requirement that
\begin{equation}
H^{\theta} \psi^{\theta}=\psi^{\theta} H^0
\label{ham1}
\end{equation}
In this case the solution of (\ref{ham1}) is $\psi^{\theta}=\exp [i\theta
\tilde{\phi}]$.  After relabeling $p=\theta +m$ where $p=0, \pm 1, \pm 2,...$ we have the solution
for the field $\tilde{\phi}$
\begin{eqnarray}
\psi = \exp [ip\tilde{\phi}] & & E^{\theta} = \frac{1}{2}(p-\theta )^2
\end{eqnarray}
We note that in the case $\theta=0$ the wavefunctional immediately
reduces to the correct form and that $m$ becomes an integer.

Re-writing in terms of $\phi$, the form of the Hamilton and, due to
its periodicity, the wavefunctional are unchanged.  We obtain
the solution $\psi = \exp [ip\phi]$ with energy
$E^{\theta} = \frac{1}{2}(p-\theta )^2$.

We note that $\psi$ is invariant under $\phi \rightarrow \phi + 2\pi$.
It is clear from the $\theta$ dependence of the energy that although
each energy level is shifted, the entire energy spectrum of the
theory is invariant under $\theta \rightarrow \theta +1$.  We note
also that the ground state is not simply
the minimum of one quadratic but a continuous function linking
the minima of the quadratics centred at $\phi = (2n+1)\pi, n=0, \pm1, \pm2,..$.
Although the separate consideration here of the field at zero momentum
resulted in no modification of the form of the vacuum wavefunctional,
the extra detail is included because the analogous procedure in section 4
does result in a modified form of the wavefunctional initially deduced
for the scalar field of zero vacuum expectation value.

We propose the same formalism for more complex theories; an explicitly
gauge invariant but $\theta$ dependent Hamiltonian with the solution
$\psi =\psi^{\theta}\psi^0$, where $\psi^{\theta}$ contains all the
$\theta$ dependence of the wavefunctional,
$H^{\theta}\psi^{\theta}=\psi^{\theta} H^0$ and $\psi$ is
gauge invariant.  $\psi^{\theta}$ and $\psi^0$ cannot be individually
gauge invariant without making the theory trivially independent of
$\theta$, which is well known not to be the general case (e.g. QCD).

\newsection{The Lagrangian and Hamiltonian of QED$_3$ with a $\theta$ term.}

In the next two subsections we shall show how the Lagrangian with a
$\theta$ term for a $2+1$ U$(1)$ theory can be obtained from breaking
the internal symmetry of the SU$(2)$ 't Hooft-Polyakov monopole and from
the compactification of a $3+1$ U$(1)$ theory.  In the third
subsection we shall discuss the
Hamiltonian of the system.  But we shall first give our motivation for
extending the model to include a scalar field.  We follow the
motivation for the topological charge
in \cite{verg} to write a $\theta$ term in the Lagrangian for purely
gauge field QED$_3$.
\begin{equation}
L=-\frac{1}{4}F_{\mu \nu}^2+\theta g \epsilon_{\mu \nu \lambda}
\partial_\mu F_{\nu \lambda}
\label{l1}
\end{equation}
In the absence of any monopoles, the $\theta$ term in this Lagrangian
is identically zero.  The gauge group of the theory is compact,
however, and the $\theta$ term
is non-zero due to the non-commutability of derivatives acting upon
the singular component of the vector field.  In other words, if we
define $F_{\mu \nu} =\partial_{\mu} A_{\nu} - \partial_{\nu} A_{\mu}$,
it is clear that in the continuum limit $\epsilon_{\mu \nu \lambda}
\partial_{\mu} F^{\nu \lambda} = 0$ and that monopoles are singular configurations for which it
is hard to write down any continuum dynamics in the Hamiltonian formalism. 

As usual the $\theta$ term may be treated as a total derivative and
written as
\begin{equation}
\theta g \epsilon_{\mu \nu \lambda} \int dS_{\mu}F_{\nu \lambda}
\end{equation}
Through the Hamiltonian formalism we therefore obtain
\begin{equation}
H=\frac{1}{2}\int d^2 x[(E_i+\theta g \epsilon_{ij}\delta (S_j-x))^2+B^2]
\label{1}
\end{equation}
where
\begin{eqnarray}
E_i &=& -i\frac{\delta}{\delta A_i}\\ \nonumber
B &=& \epsilon_{ij} \partial_{i} A_j
\label{2}
\end{eqnarray}
Hence the Hamiltonian is unchanged and the theory can be solved except
at the limit of spatial infinity.  We
shall say no more than the difficulties of treating the $\theta$
term other than as a total derivative with the immediate limitations
suggest that there is an absence of a physical field upon
which the derivative can act.  This merely hints that the theory is
incomplete as it is.

We shall quickly show that this theory is independent of $\theta$
except at spatial infinity.  Given this limitation we therefore ignore
the $\delta$-function and write (\ref{1}) as 
\begin{equation}
H=\frac{1}{2}\int d^2 x[E_i^2+b^2]
\end{equation}
This is identical to the Hamiltonian of \cite{var}.  $B$ is replaced by $b$
to ensure invariance of the Hamiltonian under large gauge
transformations as explained in section 3.3.

We require the action of a vortex creation operator upon the trial
wavefunctional to be
\begin{equation}
V \Psi = \exp [i \theta] \Psi
\label{A}
\end{equation}
From \cite{var} we know
\begin{eqnarray}
V(x)B(y) &=& B(y)V(x) + 2\pi \delta^2(x-y)V(x)\\ \nonumber
\Psi^0 &=& \int D\chi \exp [-\frac{1}{2g^2}A_i^{\chi} G^{-1} A_i^{\chi}]\\
\nonumber 
V \Psi^0 &=& \Psi^0
\end{eqnarray}
where $B=\epsilon_{ij} \partial_i A_j$.  (\ref{A}) is therefore satisfied by
\begin{equation}
\Psi = \exp[\frac{i\theta}{2\pi} \oint_C dl_i \epsilon_{ij} A_j] \int D\chi
\exp[-\frac{1}{2g^2} A_i^{\chi} G^{-1} A_i^{\chi}]
\end{equation}
The contour of the integral in the $\theta$ dependent phase is taken
about the spatial plane and so all $\theta$ dependence is seen to
reside only at spatial infinity.

\subsection{Breaking the internal symmetry of the 't Hooft-Polyakov monopole.}

It may seem a step in the wrong direction but
increasing the complexity of the theory by including a scalar field
resolves this problem.  The SU(2) 't Hooft-Polyakov monopole has been
extensively studied (for a comprehensive treatment see an excellent
new textbook \cite{wein}) and so we shall only discuss the
salient points.  The Lagrangian for the SU(2) non-abelian vector and scalar
fields is
\begin{eqnarray}
L &=& -\frac{1}{4g^2} F_{\mu \nu}^a F_{\mu \nu}^a - \frac{1}{2} D_{\mu}
\phi^{a} D_{\mu} \phi^{a} - V(\phi^{a} \phi^{a})\\ \nonumber
F_{\mu \nu}^a &\equiv& \partial_{\mu} A_{\nu}^a - \partial_{\nu}
A_{\mu}^a + \epsilon^{abc} A_{\mu}^b A_{\nu}^c\\ \nonumber
D_{\mu} \phi^a &\equiv& \partial_{\mu} \phi^a + \epsilon^{abc} A_{\mu}^b
\phi^{c}\\ \nonumber
V(\phi^a \phi^a) &\equiv& \frac{\lambda}{2}(\phi^a \phi^a - \eta^2)^2
\label{V1}
\end{eqnarray}
The symmetry of the theory can be broken from SU(2) to U(1) by
choosing $\phi^a$ to point in a specific direction in isospace
(e.g. $\phi^a =\phi^3$).  After breaking the symmetry, there is
one neutral vector field parallel to the direction of $\phi^a$ (the
photon) and
two charged vector fields orthogonal to $\phi^a$ in isospace
(the W$^{\pm}$ bosons).  We shall consider the low energy
spectrum only and hence the Lagrangian of the photon and the component
of the scalar triplet in the chosen direction of isospace.  This
scalar field has a mass proportional to $\lambda^{\frac{1}{2}}$ and a
non-zero vacuum expectation value.

To define the $\theta$ term we shall consider the gauge invariant
tensor $F_{\mu \nu}$ introduced by 't Hooft.
\begin{equation}
F_{\mu \nu} \equiv F_{\mu \nu}^a \hat{\phi}^a - \frac{1}{g}
\epsilon^{abc} \hat{\phi}^a D_{\mu} \hat{\phi}^b D_{\nu} \hat{\phi}^c
\label{6}
\end{equation}
Upon the breaking of $SU(2) \rightarrow U(1)$, by choosing $\phi^a =\phi^3$, $F_{\mu \nu}$ reduces to
give the tensor of electromagnetism, $F_{\mu \nu}^3
= \partial_{\mu} A_{\nu}^3 -\partial_{\nu} A_{\mu}^3$.  The
interpretation of $F_{\mu \nu}$ is of magnetic flux
density.  Therefore the magnetic charge of the unbroken theory, which
is topologically invariant and proportional to the topological charge
or winding
number, is given by
\begin{equation}
q = \frac{1}{8 \pi} \int d^3 x \epsilon_{\mu \nu \lambda} \partial_{\mu}
F_{\nu \lambda}
\label{7}
\end{equation}
The essential part of the $\theta$ term in this theory is therefore
$q$.  All constants may be absorbed into $\theta$ as it
is a freely varying parameter.  Using $\epsilon_{\mu \nu \lambda}
\hat{\phi}^a \partial_{\mu} F_{\nu \lambda}^a=0$, we can therefore
write the $\theta$ term of the $U(1)$ theory as
\begin{equation}
\frac{\theta}{\eta} \epsilon_{\mu \nu
\lambda} \partial_{\mu} \phi^3 F_{\nu \lambda}^3
\end{equation}
where the derivative only acts upon the scalar field and $\eta$ is the
value of the scalar field for which the potential
is a minimum.  The scalar field must
behave such that at spatial infinity $\phi^3 = \eta$ but within that
limit it can fluctuate.  This is exactly the topological term
suggested by Affleck, Harvey and Witten, \cite{Affleck}, with the
internal symmetry broken from SU(2) to U(1).

We finally propose the low energy $U(1)$ Lagrangian with a $\theta$ term
to be
\begin{equation}
L=-\frac{1}{4g^2} (F_{\mu \nu}^3)^2 - \frac {1}{2}(\partial_{\mu}
\phi^3)^2 - \frac{\lambda}{2}((\phi^3)^2-\eta^2)^2 -
\frac{\theta}{\eta} \epsilon_{\mu \nu \lambda} \partial_{\mu} \phi^3
F_{\nu \lambda}^3
\label{Lfin}
\end{equation}
In principle, one can make the scalar field heavy so it will not
affect the low-energy dynamics and only appear in the $\theta$ term of
the theory.  Let us also note that in the BPS limit $(\lambda =0)$ of
the original SU($2$) theory the scalar field will be massless.

\subsection{Compactification of U$(1)$ theory from $3+1$ to $2+1$ dimensions.}

We shall show that in the BPS limit $(\lambda =0)$ exactly the same
form of $L^{\theta}$ as above is
obtained by compactifying a U$(1)$ theory with a $\theta$ term from
$3+1$ to $2+1$ dimensions.  In this case there is no explicit
consideration of a scalar potential but the vacuum expectation value
of the scalar field is taken to be $\eta \neq 0$.  The action in $3+1$
dimensions is
\begin{equation}
S = \int d^4 x [-\frac{1}{4e^2}F_{\mu \nu}^2 -\frac{1}{4} \theta \epsilon_{\mu \nu
\lambda \rho} F_{\mu \nu} F_{\lambda \rho}]
\end{equation}
where summation over repeated indices is over
$\mu,\nu,\lambda,\rho=0,1,2,3$.  $\theta$ and $e$ are dimensionless
parameters.

Compactification of the third spatial dimension yields
\begin{equation}
S=\int d^3 x dR [-\frac{1}{4e^2}F_{\mu \nu}^2
-\frac{1}{2e^2}(\partial_{\mu} A_3)^2  - \theta \epsilon_{\mu \nu \lambda}
\partial_{\mu} A_3 F_{\nu \lambda}]
\end{equation} 
where now summation of repeated indices is over $\mu,\nu,\lambda=0,1,2$.
$R$ is the radius of compactification and the $2+1$ theory is obtained
in the limit of small R.  The dependence of $A_{\mu}$ and $A_3$ upon
the compactified coordinates, $x_3$, can be gauged away giving, after
integration,
\begin{equation}
S = \int d^3 x [-\frac{1}{4g^2}F_{\mu
\nu}^2-\frac{{\eta}^2}{2}(\partial_{\mu} \phi)^2 - \theta
\epsilon_{\lambda \mu \nu} \partial_{\lambda}
\phi F_{\mu \nu}]
\end{equation}
where $g^2=e^2 \Lambda, {\eta}^2=\frac{\Lambda}{e^2}, \Lambda \phi =
A_3, \Lambda=\frac{1}{R}$ and $\Lambda$ is the UV cut-off.  The origin
of the mass dimensions of the coupling constants in the $2+1$ U$(1)$
theory is now clear.  Given the
transformation $\phi^3=\eta (1+\phi)$, the Lagrangian above is easily
shown to be the same as (\ref{Lfin}) in the BPS limit.

\subsection{The Hamiltonian.}

The Hamiltonian is therefore
\begin{eqnarray}
H^{\theta} &=& H_A^{\theta} + H_{\phi^3}^{\theta}\\ \nonumber
H_A^{\theta} &=& \frac{1}{2} \int d^2x [g^2 (E_{A_{i}}+
\frac{2\theta}{\eta} \epsilon_{ji} \partial_j
\phi^3)^2 +\frac{1}{g^2} b^2]\\ \nonumber
H_{\phi^3} &=& \frac{1}{2}\int d^2 x
[(\pi_{\phi^3}-2\frac{\theta}{\eta}b)^2 +(\partial_i \phi^3)^2
+\eta^2 \lambda (\phi^3)^2]
\end{eqnarray}
where terms of $O(\phi^3)$ and greater have been omitted because we
are interested only in the limit of small $\lambda$.
The Hamiltonian for the scalar sector can be written in terms of the field
$\phi$,
\begin{equation}
H_{\phi}^{\theta} = \frac{1}{2} \int d^2x [
\frac{1}{\eta^2}(\pi_{\phi} - 2 \theta b)^2 +{\eta}^2 (\partial_i
\phi)^2 + 4 \eta^4 \lambda \phi^2]
\label{11}
\end{equation}
but information about the physical field, $\phi^3$, at zero momentum
is lost.

As in the case of the simple one dimensional model of section 2, the
periodicity of $\theta$ is not explicit through the Hamiltonian formalism.
In both the one dimensional case and in QED$_3$, $\theta$ is easily seen to
be periodic in the path integral formalism with the fundamental
domains of $-\frac{1}{2} < \theta < \frac{1}{2}$ for the one
dimensional model and $\frac{-1}{2q} < \theta < \frac{1}{2q}$ for
QED$_3$.  In the
one dimensional case it became apparent that, under the transformation
$\theta \rightarrow \theta +1$, it is only the entire spectrum of
solutions which
is invariant and each separate energy level is not.  But in our model
of QED$_3$ we
are considering only the low energy Hamiltonian with an ansatz
only for the vacuum or ground state of the system.  Therefore, we do
not expect this solution, but rather the entire energy spectrum, to be invariant under the transformation
$\theta \rightarrow \theta +\frac{1}{q}$ and we restrict $\theta$ to its
fundamental domain throughout the rest of this calculation.

In the compact theory point-like vortices with quantized magnetic flux
$2 \pi n$ cannot be detected by any measurement.  Within the
Hamiltonian formalism, this means that the creation operator of a
point-like vortex must be indistinguishable from the unit operator.
The operator V(x) creates such a vortex; it generates a large transformation
which belongs to the compact gauge group and so must therefore act
trivially upon all physical states.
\begin{equation}
V(x)=\exp\left\{i \int d^2y\frac{\epsilon_{ij}(x-y)_j}{(x-y)^2}
E_i(y)\right\}
\label{vortex}
\end{equation}
We therefore write $b$ in the Hamiltonian to ensure its invariance
under the action of $V(x)$ where $b$ is the singlet
part of $B$ and $B=\epsilon_{ij} \partial_{i} A_{j}$, as in \cite{var}.
So, if P is the projection operator upon the whole compact gauge
group, we can write formally $b^2=PB^2 P$.  $B$ does not commute with
V(x) but $b$ does.

Gauss' law will also be satisfied by these operators.  It is given by
\begin{equation}
\exp\{i\int d^2x\partial_i\lambda(x)E_i(x)\}|\Psi>=|\Psi>
\label{constr}
\end{equation}
where $\lambda$ here is a regular function.

\newsection{The variational ansatz for the vacuum wavefunctional.}

In this paper we require the vacuum wavefunctional in order to
calculate the energy, or vacuum expectation value, of the
Hamiltonian.  Through the functional variational technique, we
calculate the expectation value of the energy with our ansatz for the
wavefunctional and minimize this with respect to the propagators to
find the forms of the masses, propagators and the vacuum energy of the
theory.  Hence, the form of our initial ansatz is of vital importance.

We work in analogy with section $2$ but adopt a slightly modified
condition for $\Psi^{\theta}$.  We construct our wavefunctional
to be gauge invariant and have
non-trivial $\theta$ dependence in the following way.
$\Psi^{\theta}[A_i^n, \phi^3]$ contains all the $\theta$ dependence of
$\Psi [A_i,\phi^3]=\Sigma_n \Psi^{\theta} [A_i^n,\phi^3] \Psi^0
[A_i^n,\phi^3]$, where the sum over n is the sum over
large gauge transformations.  $\Psi [A_i,\phi^3]$ is the solution of
the equation
\begin{equation}
H^{\theta} \Psi [A_i,\phi^3]=E^{\theta} \Psi[A_i,\phi^3]
\label{13}
\end{equation}
such that
\begin{eqnarray}
E^{\theta} &=& \int DA_i D\phi^3 \sum_{n',n''}\Psi^{0*}[A_i^{n'},\phi^3]
\Psi^{\theta *}[A_i^{n'},\phi^3] H^{\theta}
\Psi^{\theta}[A_i^{n''},\phi^3] \Psi^0[A_i^{n''},\phi^3]\\ \nonumber
&=&  \int DA_i D\phi^3 \sum_{n',n''} \Psi^{\theta
*}[A_i^{n'} ,\phi^3]\Psi^{\theta}[A_i^{n''} ,\phi^3]\Psi^{0*}[A_i^{n'},\phi^3] H^0
\Psi^0[A_i^{n''},\phi^3]
\label{14}
\end{eqnarray}
where $H^0$ corresponds to the Hamiltonian of the system with
$\theta =0$.  Since the components of $\Psi$ cannot be individually gauge
invariant without making the theory trivially $\theta$ independent,
we have written explicitly the summation over large gauge
transformations.  We note here that each sector of the $\theta=0$
energy picks up a $\theta$ dependent phase factor.  From section 3.3
we know the
form of $H^{\theta}$ and $H^0$.  We shall first of all write the
wavefunctional in terms of $\phi$ and then transform it to depend upon
$\phi^3$, where $\phi^3=\eta(1+\phi)$, which can be considered as just
a modification of the field at zero momentum.  From a previous
calculation, \cite{var},
an ansatz is known for the gauge sector of the theory with
$\theta=0$ and which,
within the variational framework, can be used to reproduce all the
known results of dynamical mass generation, Polyakov scaling and
non-zero string tension.  $H^0_{\phi}$ is just the Hamiltonian of a
free, massive scalar field which has the solution $\exp
[-\frac{\eta^2}{2}\int \frac{d^2
k}{(2\pi)^2}(k^2+m^2)^{\frac{1}{2}}\phi (k) \phi(-k)]$.  The
combination of these two Gaussian factors gives the wavefunctional 
\begin{equation}
\Psi^0 [A_i,\phi]=\int D\chi \exp [-\frac{1}{2g^2} (A_i-\partial_i
\chi) G^{-1} (A_i -\partial_i \chi)-\frac{\eta^2}{2} \phi K^{-1}
\phi]
\end{equation}
$G$ and $K$ are, respectively, the propagators of the vector and scalar fields.  They
are parameters of the functional variational technique and so have no
explicit form at this stage.  (\ref{14}) is satisfied by 
\begin{eqnarray}
\Psi^{\theta}[A_i^n,\phi] &=& \int D\tilde{\chi}\exp [2i \theta \epsilon_{ji}
 \phi \partial_j (A_i-\partial_i \chi)]\\ \nonumber
&=& \exp [2i \theta \epsilon_{ji} \phi
\partial_j (A_i-\partial_i \chi_{\nu})]
\label{17}
\end{eqnarray}

The phase function $\chi(x)$ is parametrised as
\begin{equation} 
\chi(x)=\tilde{\chi}(x)+\chi_{\nu}(x)
\label{18}
\end{equation}
$\tilde{\chi}$ is a smooth function and $\chi_{\nu}(x)$ contains all the
discontinuities and can be written as 
\begin{equation}
\chi_{\nu}=\sum_{\alpha=1}^{n_+} \theta (x-x_{\alpha}) -
\sum_{\beta=1}^{n_-} \theta (x-x_{\beta})
\label{19}
\end{equation}
where $\theta(x-x_{\alpha})$ is a polar angle on a plane centred at
$x_{\alpha}$.  The functional measure can be written as
\begin{equation}
\int D\chi = \int
D\tilde\chi\sum_{n_+=0}^{\infty}\sum_{n_-=0}^{\infty}\frac{1}{n_+!n_-!}
\prod_{\alpha =1}^{n_{+}}\prod_{\beta=1}^{n_{-}}
\int d^2x_\alpha d^2x_\beta \Lambda^4
\label{measure}
\end{equation}
with the explicit UV momentum cut-off $\Lambda$.

We adopt the following notation for convenience;
\begin{equation}
A_i^{\chi}=A_i(x)- \partial_i \chi (x)
\label{20}
\end{equation}
and for a matrix $M(x-y)$
\begin{equation}
A_i M A_i = \int d^2 x d^2 y A_i(x)M(x-y)A_i(y)
\label{21}
\end{equation}
Using this notation, we can write down a gauge invariant ansatz for
the vacuum wavefunctional in terms of the field
$\phi$, with non-trivial $\theta$ dependence, which satisfies the above formalism.
\begin{equation}
\Psi [A_i,\phi]=\int D\chi_{\nu} D\tilde{\chi} \exp [-2i \theta \phi
\epsilon_{ij} \partial_i A_j^{\chi_{\nu}}] \exp
[-\frac{1}{2g^2} A_i^{\chi}G^{-1}A_i^{\chi}-\frac{\eta^2}{2} \phi K^{-1}
\phi]
\label{22}
\end{equation}
We now need to modify this wavefunctional to write it in terms of
the field $\phi^3$, to ensure that the extra information about the
field of zero momentum is not lost.  We write
\begin{equation}
\Psi [A_i,\phi^3] =\int D\chi \exp[\frac{2i\theta}{\eta}(\phi^3-\eta)
\epsilon_{ji} \partial_j A_i^{\chi_{\nu}}] \exp
[-\frac{1}{2g^2} A_i^{\chi}G^{-1}A_i^{\chi}
-\frac{1}{2}(\phi^3-\eta)K^{-1}(\phi^3 -\eta)]
\label{psi}
\end{equation}
It can be seen clearly from
consideration of the vacuum expectation of the Hamiltonian above that
this ansatz reduces to the
desired form in the case of $\theta=0$ and will reproduce all the
results of \cite{var}.

\newsection{The variational calculation.}

The expectation value of any operator $O(A_i,\phi)$ in the wavefunctional
(\ref{psi}) is
\begin{eqnarray}
<O(A_i,\phi)>= &Z^{-1}& \int DA_i D\phi^3 D\tilde{\chi}' D \chi_{\nu}'
D\tilde{\chi}'' D\chi_{\nu}'' \\ \nonumber
&\exp& [-\frac{2i\theta}{\eta} (\phi^3-\eta) \epsilon_{ij} \partial_i
(A_j - \partial_j \chi_{\nu}')]\\ \nonumber
&\exp& [-\frac{1}{2g^2}A_i^{\chi'} G^{-1} A_i^{\chi'} - \frac{1}{2}
(\phi^3-\eta) K^{-1} (\phi^3-\eta) ] O(A_i,\phi)\\ \nonumber
&\exp& [\frac{2i\theta}{\eta} (\phi^3-\eta) \epsilon_{ij} \partial_i (A_j -
\partial_j \chi_{\nu}'')] \\ \nonumber
&\exp& [-\frac{1}{2g^2}A_i^{\chi''} G^{-1}
A_i^{\chi''} - \frac{1}{2} (\phi^3-\eta) K^{-1} (\phi^3-\eta) ]
\label{expt1}
\end{eqnarray}
If $O(A_i)$ is explicitly gauge invariant we may shift the integration
variable $A_i^{\chi''} \rightarrow A_i$.  With the redefinition $\chi
=\chi' -\chi''$ and $\zeta = \chi' +\chi''$, and similarly for
$\tilde{\chi}$ and $\chi_{\nu}$, the expectation value reduces to 
\begin{eqnarray}
<O(A_i,\phi)>= &Z^{-1}& \int DA_i D\phi^3 D\tilde{\zeta} D\tilde{\chi}
D\zeta_{\nu} D\chi_{\nu} \\ \nonumber
&\exp& [-\frac{2i\theta}{\eta} (\phi^3-\eta) \epsilon_{ij} \partial_i
(A_j - \partial_j \chi_{\nu})]\\ \nonumber
&\exp& [-\frac{1}{2g^2}A_i^{\chi} G^{-1} A_i^{\chi} - \frac{1}{2}
(\phi^3-\eta) K^{-1} (\phi^3-\eta) ] O(A_i,\phi)\\ \nonumber
&\exp& [\frac{2i\theta}{\eta} (\phi^3-\eta) \epsilon_{ij} \partial_i
A_j] \\ \nonumber
&\exp& [-\frac{1}{2g^2}A_i G^{-1} A_i - \frac{1}{2} (\phi^3-\eta) K^{-1}
(\phi^3-\eta) ]
\label{24}
\end{eqnarray}
The integral over $D\zeta = D\tilde{\zeta} D\zeta_{\nu}$ just gives the
volume of the gauge group and so cancels with the denominator.

\subsection{Calculation of the energy density.}

First we shall evaluate $Z$.
\begin{eqnarray}
Z &=& \int DA_i D\phi^3 D\tilde{\chi} D\chi_{\nu} \exp [\frac{2i \theta}{\eta}
(\phi^3-\eta) \epsilon_{ij} \partial_i \partial_j \chi_{\nu}-\frac{1}{g^2} A_i
G^{-1} A_i \\ \nonumber
&-& (\phi^3-\eta) K^{-1} (\phi^3-\eta)
+\frac{1}{g^2} \partial_i \chi G^{-1} A_i
-\frac{1}{2g^2} \partial_i \chi G^{-1}\partial_i \chi]
\label{25}
\end{eqnarray}
By completing the squares with the two changes of variable,
\begin{eqnarray}
\phi^3 &\rightarrow& \phi'^3 = \phi^3 - \frac{i\theta}{\eta}
\epsilon_{ij} \partial_i \partial_j \chi_{\nu} K = \phi^3 -\frac{2i\pi
\theta}{{\eta}}  \rho K\\ \nonumber
A_i &\rightarrow& A_i' = A_i - \frac{1}{2} \partial_i \chi
\label{26}
\end{eqnarray}
and omitting the dashes on the new variables, one obtains
\begin{eqnarray}
Z &=& Z_a Z_{\phi^3} Z_{\chi} Z_{\nu}\\ \nonumber
Z_a &=& \int DA_i \exp [-\frac{1}{g^2}A_i G^{-1} A_i] = det[g^2 \pi G]
\\ \nonumber
Z_{\phi^3} &=& \int D\phi^3 \exp [- (\phi^3-\eta) K^{-1} (\phi^3-\eta)] =
det[\pi K]^{\frac{1}{2}}\\ \nonumber
Z_{\chi} &=& \int D\tilde{\chi} \exp [-\frac{1}{4g^2}\partial_i
\tilde{\chi} G^{-1} \partial_i \tilde{\chi}] = det[4g^2 \pi
\frac{1}{\partial^2} G]^{\frac{1}{2}}\\ \nonumber
Z_{\nu} &=& \int D\chi_{\nu} \exp [-\frac{1}{4g^2}\partial_i
\chi_{\nu} G^{-1} \partial_i \chi_{\nu} -\frac{{\theta}^2}{{\eta}^2} 
\epsilon_{ij} \partial_i \partial_j \chi_{\nu} K \epsilon_{ij}
\partial_i \partial_j \chi_{\nu}]
\label{27}
\end{eqnarray}

Details of the derivative transformation - used here and in evaluation of the
following Gaussian integrals over the singular function $\chi_{\nu}$ -
that establishes the connection between derivatives of $\chi_{\nu}$
and the distribution function of its singularities or
vortices, $\rho$, are given in the appendix.  Any singularities in
$\phi (x)$ remaining after the change of
variables are taken to contribute an infinite action and so are
ignored.  The singularities in $\chi_{\nu}$, however, cannot be
ignored.  

To evaluate $Z_{\nu}$ we shall write it as a
partition function of a gas of vortices and use the standard trick of
\cite{polyakov}, \cite{samuel}.
\begin{eqnarray}
Z_v&=&\sum_{n_+,n_- = 0}^{\infty}
\prod_{\alpha=1}^{n_{+}}\prod_{\beta=1}^{n_{-}}
\int d^2x_\alpha d^2x_\beta z^{n_++n_-} \\
&\exp&\left\{-\frac{1}{4g^2}[\sum_{\alpha,\alpha'}D(x_\alpha-x_{\alpha'})
+\sum_{\beta,\beta'}D(x_\beta-x_{\beta'})-
\sum_{\alpha,\beta}D(x_\alpha-x_{\beta})]\right\} \nonumber
\label{part}
\end{eqnarray}
where the vortex - vortex interaction potential $D(x)$ and the vortex
fugacity $z$ are given by
\begin{eqnarray}
D(x)&=&8\pi^2\int \frac{d^{2}k}{(2\pi)^{2}}
(k^{-2} G^{-1}(k)+\frac{4\theta^2 g^2}{\eta^2} K(k))\cos(kx)\\ \nonumber
z&=&\Lambda^2\exp\{-\frac{1}{8g^2}D(0)\}
\label{potential}
\end{eqnarray}
We expect the UV behaviour of $G(k)$ and $K(k)$ at large momentum to
be the same as in
the free theory ($G(k)\rightarrow k^{-1},K(k)\rightarrow k^{-1}$).
The vortex fugacity is the smallest variable in the theory,
$z<<g^2<<\Lambda$, where in the limit of weak coupling,
\begin{equation}
z=\Lambda^2\exp\{-\frac{\pi}{2}\frac{\Lambda}{g^2}(1+\frac{4
\theta^2 g^2}{\eta^2})\}=\Lambda^2\exp\{-\frac{\pi}{2}(\frac{\Lambda}{g^2}+\frac{4\theta^2 g^2}{\Lambda})\}
\label{zpar}
\end{equation}
where we have identified $\Lambda= g \eta$ from the compactification
of the U$(1)$ theory from $3+1 \rightarrow 2+1$ dimensions or,
alternatively, from the masses of the charged
vector bosons of the theory.  

We will need to calculate correlation functions of the vortex density
and so, following \cite{var}, we write the vortex density as 
\begin{equation}
\rho(x)=\sum_{\alpha,\beta}\delta(x-x_\alpha)-\delta(x-x_\beta)
\end{equation}
Introducing a source term the exponential factor including the vortex
fugacity in (\ref{part}) can
be written as
\begin{equation}
\Lambda^{2(n_++n_-)}\int D\chi\exp\{-2g^2\chi
D^{-1}\chi+i\rho\chi+i\rho J\}
\end{equation}
and the sum over the number of vortices and anti-vortices gives
\begin{equation}
Z_v=\int D\chi\exp\{-2g^2(\chi -J) D^{-1} (\chi -J)+
\int _x2\Lambda^2\cos\chi(x)\}\mid_{J=0}
\label{sine}
\end{equation}
Calculating the functional derivatives with respect to the source term
yields
\begin{equation}
<\rho(x)\rho(y)>=4g^2D^{-1}(x-y)-16g^4<D^{-1}\chi(x)D^{-1}\chi(y)>
\end{equation}
The propagator of $\chi$ is easily calculated.  First the cosine
potential is rewritten in the normal ordered form.
\begin{equation}
\cos \chi = \frac{z}{\Lambda^2} :\cos \chi:
\label{norm}
\end{equation}
Therefore, to first order in $z$, the propagator of $\chi$ is
\begin{equation}
\int d^2 x e^{ikx}<\chi(x)\chi(0)>=
\frac{1}{4g^2D^{-1}(k)+2z}=\frac{D(k)}{4g^2}
-z\frac{D^2(k)}{8g^4} + o(z^2)
\end{equation}
To first order, the correlator of the vortex densities is then
\begin{equation}
C(k)=\int d^2x e^{ikx}<\rho(x)  \rho(0)>=2z + o(z^2)
\label{rhoc}
\end{equation}
as in \cite{var} but with a $\theta$ dependent modification to $z$.

Now we can calculate the expectation value of the Hamiltonian.
\begin{eqnarray}
<H^{\theta}> &=& <H_A^{\theta}> + <H_{\phi^3}^{\theta}>\\ \nonumber
&=& Z^{-1}\int DA_i D\phi D\tilde{\chi} D\chi_{\nu} \exp
[\frac{2i\theta}{\eta} (\phi^3-\eta) \epsilon_{ij}\partial_i \partial_j
\chi_{\nu}]\\ \nonumber
&\exp& [-\frac{1}{2g^2}A_i^{\chi}G^{-1}A_i^{\chi}-\frac{1}{2}(\phi^3-\eta)
K^{-1}(\phi^3-\eta)][H_A^0+H_{\phi^3}^0]\\ \nonumber
&\exp& [-\frac{1}{2g^2}A_iG^{-1}A_i-\frac{1}{2}(\phi^3-\eta)
K^{-1}(\phi^3-\eta)]
\label{30}
\end{eqnarray}
First we shall consider the purely gauge field sector.
\begin{eqnarray}
&<& \frac{g^2}{2} \int d^2 x (E_{A_{i}} +2\theta
\epsilon_{ji}\partial_{j}(\phi^3-\eta))^2> = \frac{g^2}{2}
Z^{-1} \int DA_i D\phi^3 D\tilde{\chi} D\chi_{\nu} \\ \nonumber
&[& \frac{2}{g^2} TrG^{-1} - \frac{1}{g^2}A_i G^{-2} A_i] \exp
[\frac{2i\theta}{\eta} (\phi^3-\eta)
\epsilon_{ij}\partial_i \partial_j \chi_{\nu} -\frac{1}{g^2} A_i
G^{-1}A_i\\ \nonumber
&-& (\phi^3-\eta) K^{-1} (\phi^3-\eta)
+\frac{1}{g^2}\partial_i \chi G^{-1}A_i -
\frac{1}{2g^2} \partial_i \chi G^{-1} \partial_i \chi]
\end{eqnarray}
Completing the squares and performing the functional integrals gives
\begin{eqnarray}
\frac{1}{V}<\frac{g^2}{2} \int d^2 x(E_{A_{i}}+2\theta
\epsilon_{ji}\partial_{j}(\phi^3-\eta))^2> &=& \frac{1}{2} \int \frac{d^2
k}{(2\pi)^2}[\frac{1}{2} G^{-1}(k) -
\frac{\pi^2}{g^2}  k^{-2} C(k) G^{-2}]\\
\nonumber
&=& \frac{1}{4}\int \frac{d^2 k}{(2\pi)^2} [G^{-1}(k)
-\frac{4\pi^2}{g^2} z k^{-2} G^{-2}(k)]
\label{31}
\end{eqnarray}
The Gaussian integral over $\chi_{\nu}$ is transformed into a
correlation function of $\rho$.  This procedure is given in more
detail in the appendix.

Now for the magnetic term.  In every gauge invariant
state $<b^2>=<B^2>$ by definition.  We will therefore calculate
$<B^2>$.  But since it
is not itself gauge invariant some care is needed with the integrals
over $\chi$ and $\zeta$.
\begin{eqnarray}
<\frac{1}{2g^2} b^2> &=& Z^{-1} \int DA D\phi^3 D\tilde{\chi}' D
\chi_{\nu}' D\tilde{\chi}'' D\chi_{\nu}'' \frac{1}{2g^2}
[\epsilon_{ij} \partial_i A_j]^2 \\ \nonumber
&\exp& [\frac{2i\theta}{\eta} (\phi^3-\eta) \epsilon_{ij}
\partial_i \partial_j (\chi_{\nu}'-\chi_{\nu}'') -\frac{1}{2g^2}
A_i^{\chi'} G^{-1} A_i^{\chi'} \\ \nonumber
&-& \frac{1}{2g^2} A_i^{\chi''} G^{-1}
A_i^{\chi''}- (\phi^3-\eta) K^{-1} (\phi^3-\eta)]\\ \nonumber 
&=& Z^{-1}\int DA_i D\phi^3 D\chi D\zeta \frac{1}{2g^2} [\epsilon_{ij} \partial_i
(A_j + \frac{1}{2}\partial_j (\zeta-\chi))]^2\\ \nonumber
&\exp& [\frac{2i\theta}{\eta} (\phi^3-\eta) \epsilon_{ij}\partial_i \partial_j
\chi_{\nu} -\frac{1}{2g^2} A_i^{\chi} G^{-1}A_i^{\chi}\\ \nonumber
&-& \frac{1}{2g^2} A_i
G^{-1}A_i -(\phi^3-\eta) K^{-1}(\phi^3-\eta)]
\label{32}
\end{eqnarray}
We have again used $\chi=\chi ' - \chi ''$ and $\zeta=\chi ' + \chi
''$.  The linear term in $\zeta$ disappears due to the symmetry of its
measure.  The term quadratic in $\zeta$ is independent of $G,K$ and
$\theta$ and so contributes nothing of interest to either the energy or
the minimization equations.  After completing the squares we obtain
\begin{eqnarray}
\frac{1}{V}<\frac{1}{2g^2}b^2> &=& Z_a^{-1}\int DA_i D\phi^3 D\chi \frac{1}{2g^2} [\epsilon_{ij}
\partial_i A_j]^2\exp [ -\frac{1}{g^2} A_i G^{-1}A_i]\\ \nonumber
&=& \frac{1}{4} \int \frac{d^2 k}{(2\pi)^2} k^2 G(k)
\label{33}
\end{eqnarray}
So, for the purely gauge field sector we obtain
\begin{equation}
\frac{1}{V}<H_A^{\theta}>=\frac{1}{4} \int \frac{d^2 k}{(2\pi)^2}
[G^{-1}(k) + k^2 G(k) -\frac{4\pi^2}{g^2} z k^{-2}G^{-2}(k)]
\label{34}
\end{equation}
This is of the same form as \cite{var} but with the modified
expression for $z$.

Following the procedure above we calculate the vacuum expectation
value of the Hamiltonian of the scalar field. 
\begin{eqnarray}
\frac{1}{V} &<& \frac{1}{2} \int d^2 x (\pi_{\phi^3}
-\frac{2\theta}{\eta}b )^2>=\frac{1}{2} Z^{-1}\int DA_i D\phi^3
D\tilde{\chi} D\chi_{\nu}\\ \nonumber
&[& K^{-1}- (\phi^3-\eta) K^{-2} (\phi^3-\eta)]
\exp [\frac{2i\theta}{\eta} (\phi^3-\eta) \epsilon_{ij}\partial_i \partial_j
\chi_{\nu} \\ \nonumber
&-& \frac{1}{g^2} A_i G^{-1}A_i-(\phi^3-\eta)
K^{-1}(\phi^3-\eta)
+ \frac{1}{g^2}\partial_i \chi G^{-1}A_i -
\frac{1}{2g^2} \partial_i \chi G^{-1}\partial_i
\chi]\\ \nonumber
&=& \int \frac{d^2 k}{(2\pi)^2}[\frac{1}{4} K^{-1}(k) +
\frac{2 \pi^2 \theta^2}{\eta^2} C(k)] \\ \nonumber
&=& \int \frac{d^2 k}{(2\pi)^2} [\frac{1}{4} K^{-1}(k)
+\frac{4\pi^2 \theta^2}{\eta^2} z]
\label{35}
\end{eqnarray}
Similarly the terms quadratic in $\phi$ give
\begin{eqnarray}
\frac{1}{V}<\frac{1}{2} \int d^2 x(\partial_i \phi^3)^2> &=&  \frac{1}{4} \int \frac{d^2
k}{(2\pi)^2}[k^2 K(k) -\frac{16 \pi^2 \theta^2}{\eta^2} z
k^2 K^2(k)]\\ \nonumber
\frac{1}{V}<\frac{1}{2} \int d^2 x \lambda \eta^2 (\phi^3-\eta)^2> &=& \frac{1}{4} \int \frac{d^2
k}{(2\pi)^2}[\lambda \eta^2 K(k) -16 \lambda \pi^2 \theta^2
z K^2(k)]
\end{eqnarray}

Therefore, for the scalar field we obtain
\begin{equation}
\frac{1}{V}<H_{\phi^3}^{\theta}>=\frac{1}{4} \int \frac{d^2
k}{(2\pi)^2}[K^{-1}(k)+\frac{16 \pi^2 \theta^2}{\eta^2}z
+(K(k)-\frac{16 \pi^2 \theta^2}{\eta^2}zK^2(k))(k^2+\lambda
\eta^2)] 
\label{36}
\end{equation}

\subsection{Minimisation of the vacuum energy density.}

Details of the functional minimisation of the energy density with
respect to the vector field propagator, $G(k)$, and the scalar field
propagator, $K(k)$, are given in appendix B.

We obtain the simple minimization equation
\begin{eqnarray}
0 &=& \frac{1}{4}[k^2-G^{-2}(k)]-\frac{\pi^4}{g^4}k^{-2}G^{-2}(k)z\int
\frac{d^2p}{(2\pi)^2}p^{-2}G^{-2}(p)\\ \nonumber
 &+& \frac{4\theta^2 \pi^4}{\eta^2
g^2}k^{-2}G^{-2}(k)z \int \frac{d^2p}{(2\pi)^2}[1-K^2(p)(p^2+\lambda \eta^2)]
\end{eqnarray}
with the solution
\begin{eqnarray}
G^{-2}(k) &=& \frac{k^4}{k^2+m^2}\\ \nonumber
m^2 &=& \frac{4\pi^4}{g^4}z[\int \frac{d^2
p}{(2\pi)^2}p^{-2}G^{-2}(p)-\frac{4\theta^2 g^2}{\eta^2}\int
\frac{d^2p}{(2\pi)^2}[1-K^2(p)(p^2+8\lambda \eta^2)]]
\end{eqnarray}
and
\begin{equation}
0=\frac{1}{4}[\frac{16 \pi^4 \theta^2}{g^2 \eta^2}z\int
\frac{d^2p}{(2\pi)^2}[p^{-2}G^{-2}(p)-\frac{4\theta^2
g^2}{\eta^2}(1-K^2(p)(p^2+\lambda \eta^2))]+(k^2+\lambda \eta^2)-K^{-2}(k)]
\end{equation}
with the solution
\begin{eqnarray}
K^{-2}(k) &=& k^2+m_{\phi}^2+m_{\theta}^2\\ \nonumber
m_{\phi}^2 &=& \lambda \eta^2\\ \nonumber
m_{\theta}^2 &=& \frac{16\pi^4 \theta^2}{g^2 \eta^2}z\int
\frac{d^2p}{(2\pi)^2}[p^{-2}G^{-2}(p)-\frac{4\theta^2
g^2}{\eta^2}(1-K^2(p)(p^2+\lambda \eta^2))]\\ \nonumber
&=& \frac{4\theta^2 g^2}{\eta^2} m^2
\end{eqnarray}

Explicit evaluation of the photon mass gives
\begin{equation}
m^2 = \frac{\pi^3}{g^4} z[\Lambda^2 - m^2 \log (\frac{\Lambda^2 +
m^2}{m^2}) - \frac {4\theta^2 g^2}{\eta^2} (m_{\phi}^2+m_{\theta}^2)
\log (\frac{\Lambda^2 + m_{\phi}^2 +
m_{\theta}^2}{m_{\phi}^2 + m_{\theta}^2})]
\label{mass2}
\end{equation}
The $\theta$ dependence of $z$ has been shown to be sub-leading in
terms in $\Lambda$, (\ref{zpar}).  In the limit of large UV cut-off
momentum, $\Lambda$, all the $\theta$ dependence of z is therefore
suppressed in sub-leading terms in $\Lambda$.  In the BPS limit
$(\lambda =0)$, the $\theta$ dependence in (\ref{mass2}) only occurs
at $O(z^2)$ and so is suppressed further.

In the BPS limit we can write, to first order in $z$
\begin{eqnarray}
m^2 &=& \frac{\pi^3}{g^4} \Lambda^2 z\\ \nonumber
m_{\phi}^2 &=& 0\\ \nonumber
m_{\theta}^2 &=& 4\frac{\theta^2 g^2}{\eta^2} m^2
\end{eqnarray}
We should note here that $m^2$ is in agreement with \cite{var} to
first order in $z$.

\subsection{Evaluation of the vacuum energy density.}

Now we have the forms of the propagators and the masses for the fields
we can consider the $\theta$ dependence of the vacuum expectation
value of the Hamiltonian.  The vacuum energy densities of the gauge
and scalar sectors evaluated in the limit of large UV cut-off,
$\Lambda$, are,
\begin{eqnarray}
\frac{1}{V}<H_A^{\theta}> &=& \frac{1}{8\pi} [\frac{2}{3} \Lambda^3
+\frac{1}{3} m^3] -\int \frac{d^2k}{(2\pi)^2}\frac{\pi^2}{g^2} z
k^{-2} G^{-2}(k)  \\ \nonumber
\frac{1}{V}<H_{\phi^3}^{\theta}> &=& \frac{1}{8\pi} [\frac{2}{3}
\Lambda^3 +\frac{1}{3}(\lambda \eta^2 +\frac{4\theta^2
g^2}{\eta^2}m^2)^{\frac{3}{2}}+\lambda \eta^2 (\Lambda -  (\lambda
\eta^2 +\frac{4\theta^2 g^2}{\eta^2}m^2)^{\frac{1}{2}})]\\ \nonumber
&+& \int
\frac{d^2k}{(2\pi)^2} \frac{4\pi^2
\theta^2}{\eta^2}z(1-K^2(k)(k^2+\lambda \eta^2))
\end{eqnarray}
The last term in each expression combine to give the exact form of $m^2$.
It has been shown that the $\theta$ dependence of $z$ is a sub-leading
term in $\Lambda$, (\ref{zpar}).  Therefore, the gauge sector of the
theory is manifestly independent of $\theta$.  The scalar sector does
have an explicit dependence upon $\theta$ but it is always suppressed
by an order of $z$. 
It is interesting to note, however, that it is in the BPS limit
$(\lambda=0)$ and in the limit of a very massive scalar field
that the $\theta$ dependence of the total vacuum energy density is
most greatly suppressed.  In the BPS limit the total vacuum energy density is
\begin{equation}
\frac{1}{V}<H^{\theta}> = \frac{1}{8\pi}[\frac{4}{3}\Lambda^3
+\frac{1}{3}(1+ \frac{8 \theta^3 g^3}{\eta^3})m^3]
-\frac{1}{4}\frac{g^2}{\pi^2}m^2 
\end{equation}
where the $\theta$ dependence of the scalar sector is suppressed in
terms of $O(z^{\frac{3}{2}})$.  In the limit of a very massive scalar
field ($\lambda$ becomes large) the vacuum energy density is,
\begin{equation}
\frac{1}{V}<H^{\theta}> = \frac{1}{8\pi} [\frac{4}{3} \Lambda^3
+\frac{1}{3}m^3 - \frac{2}{3}(\lambda \eta^2)^{\frac{3}{2}} +\lambda
\eta^2 \Lambda] -\frac{1}{4} \frac{g^2}{\pi^2} m^2
\end{equation}
In the limit of a very massive scalar field, the only $\theta$
dependence of the scalar sector is in the modified form of $z$, which
we have already shown to be sub-leading in terms of $\Lambda$.  So in
this limit we recover the exact result of Vergeles.

\subsection{Expectation value of the Wilson loop.}

Finally, we can calculate the expectation value of the Wilson loop, as
in \cite{var}, to see how the $\theta$ dependence affects confinement.
\begin{equation}
W_C= <\exp\{il\oint_C A_{i}dx_{i}\}> = <\exp\{il\int_SBdS\}>
\end{equation}
where $l$ is an arbitrary integer and the integral is over the area $S$ 
bounded by the loop $C$.  We have written $B$ rather than $b$, since
this exponential operator is invariant under transformations $B(x)\rightarrow B(x)+2\pi$, generated by the vortex operator.
\begin{eqnarray}
W_C &=& Z^{-1} \int DA_i D\phi D\tilde{\chi} D\chi_{\nu} \exp
[\frac{2i\theta}{\eta} (\phi^3-\eta) \epsilon_{ij} \partial_i \partial_j
\chi_{\nu}\\ \nonumber
&-&\frac{1}{g^2}A_i G^{-1} A_i +\frac{1}{g^2}
\partial_i \chi G^{-1} A_i -\frac{1}{2g^2} \partial_i \chi G^{-1}
\partial_i \chi \\ \nonumber
&-& (\phi^3-\eta) K^{-1} (\phi^3-\eta) + il\int_SBdS]
\end{eqnarray}
After completing the squares
\begin{equation}
W_C = W_0 W_{\nu}
\end{equation}
where
\begin{eqnarray}
W_0 &=& Z_a^{-1} \int DA_i \exp [-\frac{1}{g^2}A_i G^{-1} A_i + il\int_SBdS]\\
\nonumber
W_{\nu} &=& Z_{\nu}^{-1}\int D\chi_{\nu}
\exp[-\frac{1}{4g^2}\partial_i\chi_{\nu}
(G^{-1}+\frac{4\theta^2 g^2}{\eta^2}\partial^2 K) \partial_i\chi_{\nu}\\
\nonumber
&-& \frac{il}{2} \int_S dS \epsilon_{ij} \partial_i \partial_j \chi_{\nu} dx_i]
\end{eqnarray}
In weak coupling $W_0$ becomes
\begin{eqnarray}
W_0 &=& \exp\left\{-\frac{l^2}{2} \int_{x,y}<B(x)B(y)>d^2x
d^2y\right\}\\ \nonumber
&=& \exp [-\frac{l^2 g^2}{4} mS]
\label{wa}
\end{eqnarray}
in the limit $k \rightarrow 0$.  This term is independent of $\theta$
and gives the string tension $\sigma = \frac{l^2 g^2}{4}m$.

$W_{\nu}$ differs from unity only for odd $l$, for which it can be calculated.
\begin{equation}
W_v=<\exp\{i\pi\int_S\rho(x)d^2x\}>=\int D\chi \exp\{-2g^2
\chi D^{-1}\chi+\int _{x}2
\Lambda^2\cos(\chi(x)-\alpha(x))\}
\label{wvort}
\end{equation}
where $\alpha (x)$ is zero outside and $\pi$ inside the loop.
Following the normal ordering prescription for a scalar field given in
subsection 4.1, and noting that the solution to the classical equations which
contributes to the leading order result is $\chi(x)=0$, we obtain the
solution $W_{\nu}=\exp [-2zS]$.  As in \cite{var} this is a sub-leading
correction to the string tension $(2z<<\sigma)$ where the $\theta$
dependence in $z$ (and hence also in the factor $z^{\frac{1}{2}}$ in m) is greatly suppressed as a sub-leading term in $\Lambda$.

\newsection{Axionic confining strings}

Polyakov showed that purely gauge field compact QED$_3$ is equivalent to a
non-standard string theory, \cite{Polyakov2}.  We shall show in this
section that our proposed Lagrangian for low energy compact QED$_3$ with a
scalar field and a $\theta$ term gives rise to the same non-standard
string theory but with a $\theta$ dependent modification of the mass
of the photon.  This modification of the mass is a sub-leading term in
the UV momentum cut-off and is as predicted in (\ref{zpar}).

We shall first give a brief revue of the relevant details from
\cite{Polyakov2}.  The Wilson loop calculated in compact QED$_3$ is
\begin{eqnarray}
W(C) &=& \int DA_{\mu} \exp [-S(A) + i\oint dx_{\mu} A_{\mu}]\\ \nonumber
S(A) &=& \frac{1}{4g^2}\int d^3x F_{\mu \nu}^2
\label{wils}
\end{eqnarray}
Here $F_{\mu \nu} = \partial_{\mu} A_{\nu} - \partial_{\nu} A_{\mu}$.
In the calculation of $W(C)$ one must include the monopole
configurations of the vector field.  As a result (\ref{wils}) has the
representation $W(C) = W_0(C) W_M(C)$ where the first factor comes from
the Gaussian integration over the vector field and the second factor
is from the contribution of the point-like monopoles.  As in the
instanton gas calculations, \cite{polyakov}, the contribution
of one monopole at point $x$ is considered first.
\begin{eqnarray}
W_M^1(x,C) &\propto& \exp (-\frac{a}{g^2}+i\eta(x,C))\\ \nonumber
\eta(x,C) &=& \oint_C dy_{\mu} A_{\mu}^{(mon)}(x-y) = \int_{\Sigma_C}
d^2 \sigma_{\mu} (y) \frac{(x-y)}{|x-y|^3}
\end{eqnarray}
$\eta (x,C)$ is the solid angle formed by the point $x$ and the
contour $C$.  $\Sigma_C$ is an arbitrary surface bounded by the
contour C.  $a=M_W \epsilon(\frac{\lambda}{g^2})$ where
$M_W=g\eta=\Lambda$ as stated in section 3.  For $\lambda =0$, $\epsilon(\frac{\lambda}{g^2})=4\pi$.  Summation over all possible monopole configurations leads
to the scalar field theory
\begin{equation}
W_M(C) \propto \int D\phi \exp [-g^2 \int d^3x(\frac{1}{2}(\partial \phi)^2
+m^2(1-\cos (\phi + \eta)))]
\label{aa}
\end{equation}
with $m^2 \propto \exp [-\frac{a}{g^2}]$.
Rewriting this theory in terms of an effective action by introducing
a rank two anti-symmetric tensor field, $B$, Polyakov suggested a new
type of strings, which he called confining strings.  Let us consider
the axionic confining strings; the strings in our theory with an extra
scalar (axionic) field coupled to the photon field in a $\theta$ term.

We shall show that the proposed low energy theory for QED$_3$ with a
scalar field and a $\theta$ term is the equivalent of (\ref{aa}) with a
modification of the photon mass, $m$.  Working in direct analogy to
the above, we calculate the Wilson loop.
\begin{eqnarray}
W(C) &=& \int DA_{\mu} \exp [-S(A,\phi^3) + i\oint dx_{\mu} A_{\mu}]
\\ \nonumber
S(A,\phi^3) &=& \int d^3x [\frac{1}{4g^2}F_{\mu \nu}^2
+\frac{1}{2}(\partial_{\mu} \phi^3)^2
+\frac{2 i \theta}{\eta}\partial_{\mu}\phi^3 \tilde{F}_{\mu}]\\
\nonumber
&=& \int d^3x [\frac{1}{4g^2}F_{\mu \nu}^2 - \frac{1}{2} \phi^3 \Box
\phi^3 -\frac{2i\theta}{\eta} \phi^3 \partial_{\mu} \tilde{F}_{\mu}]
+2i\theta q
\end{eqnarray}
where $\tilde{F}_{\mu} = \frac{1}{2}\epsilon_{\mu \nu \lambda} F_{\nu
\lambda}$ and $q=\int d^2 S_{\mu} \tilde{F}_{\mu}$.  The $\phi^3$ field is eliminated by Gaussian integration
with the transformation $\phi^3(x) \rightarrow \phi'^3(x) = \phi^3(x) +
\frac{2i\theta}{\eta} \Box^{-1}(x)\partial_{\mu} \tilde{F}_{\mu}(x)$.
Care is needed with the definition of the inverse D'Alembertian, the
action of which upon an arbitrary function $f(x)$ is, 
\begin{equation}
\Box^{-1}(x)f(x) = \int d^3x' \Box^{-1}(x-x') f(x')
\end{equation}
The action of the D'Alembertian gives the correct result allowing the
interpretation of $\Box^{-1}$ as a Green's function
\begin{equation}
\Box (x) \Box^{-1}(x-x') = \delta(x-x')
\end{equation}
Therefore, integrating out the $\phi^3$ field we obtain
\begin{equation}
S(A,\phi^3)= \int d^3x [\frac{1}{4g^2} F_{\mu
\nu}^2-\frac{2\theta^2}{\eta^2} \int d^3x'\Box^{-1}(x-x')\partial_{\mu}
\tilde{F}_{\mu}(x')\partial_{\nu}\tilde{F}_{\nu}(x)] +2i\theta q
\label{aaa}
\end{equation}
We are working in the limit of zero scalar potential here.  The
contribution of one monopole at point $x$ is therefore
\begin{equation}
W_M^1(x,C) \propto \exp (-\frac{a}{g^2}+b+i(\eta(x,C)+2\theta q))
\end{equation}
where $b$ comes from the evaluation of the $\theta^2$ term in
(\ref{aaa}), which can be written as
\begin{equation}
\frac{2\theta^2 g^2}{\Lambda^2}\int d^3x d^3x'\Box^{-1}(x-x')\partial_{\mu}
\tilde{F}_{\mu}(x') \partial_{\nu} \tilde{F}_{\nu}(x)
\label{aaaa}
\end{equation}
Immediately we see that (\ref{aaaa}) has the mass dimension of -1 in
agreement with the $\theta^2$ modification of $z$, (\ref{zpar}).  From \cite{polyakov}, we know the configuration of the A field due to
monopoles gives rise to
\begin{equation}
\tilde{F}_{\mu} = \frac{1}{2}(\frac{x_{\mu}}{|x|^3}-4\pi \delta_{\mu
3} \theta(x_3) \delta(x_1) \delta(x_2))
\label{bbb}
\end{equation}
As in \cite{polyakov}, each monopole is surrounded with a sphere of
radius $R$ such that $M_W^{-1} << R << |x_{ab}|$ where $x_{ab}$ is the
distance between two monopoles, located at $x_a$ and $x_b$.  Inside
the sphere (\ref{bbb}) is not valid and the influence of other
monopoles may be neglected.  This is the region that gives rise to the
so-called self-pseudoenergy of the monopoles, $a$.  Monopoles of
charge $>1$ are neglected as they can be considered as the limit of
two or more monopoles in close proximity and these configurations have
been shown to be inessential, \cite{polyakov}.  Only far separated
monoples are important in the infra-red region.  For large separation,
\begin{equation}
\partial_{\mu}\tilde{F}_{\mu} \simeq -2\pi \delta^3(x)
\end{equation}
Writing (\ref{aaaa}) in momentum space we obtain,
\begin{equation}
\frac{2\theta^2 g^2}{\Lambda^2}4\pi^2 \int d^3x d^3x'\int
\frac{d^3k}{(2\pi)^3} \int \frac{d^3k'}{(2\pi)^3} \int
\frac{d^3k''}{(2\pi)^3} k^{-2}\exp[i(k(x-x')+k'x'+k''x)] \propto
\frac{\theta^2 g^2}{\Lambda}
\end{equation}
This is in direct agreement with (\ref{zpar}).  0ur proposed
Lagrangian for low energy QED$_3$ hence gives an equivalent form of
(\ref{aa}) with the modifications of $m^2$ replaced by $m'^2$ and
$\eta(x,C)$ replaced by $\eta'(x,C)$ where $m'^2
= m^2 \exp[-const.\frac{\theta^2 g^2}{\Lambda}]$ and
$\eta'(x,C)=\eta(x,C)+2\theta q$.  The
modified form of the photon mass will not change the rest of the
formalism of \cite{Polyakov2}.  The constant shift in $\eta$ will have
an effect upon the monopole configuration, or shape of the surface,
$\Sigma_C$, that minimizes the action. 
Because of the integration over all $x$, which is equivalent to a sum
over all angles, such a constant shift should have no effect within the
formalism of \cite{Polyakov2}.

\newsection{Conclusion}

We have found that it is much more natural
to include a scalar Higgs field to consider a $\theta$ term in
QED$_3$.  The theory without a scalar field gives the same result of
being independent of $\theta$ but excludes the limit of spatial infinity.  From
consideration of the $\theta$ term in the non-abelian
Lagrangian of the SU(2) 't Hooft - Polyakov monopole we propose
such a term in QED$_3$ in which the gauge and scalar fields are
coupled.  The term we propose is exactly of the form of the
topological term proposed by Affleck et al, \cite{Affleck}.  We find that this term is expected if the $2+1$ dimensional
theory is considered as a result of dimensional reduction of a purely
gauge $U(1)$ theory with a $\theta$ term in $3+1$ dimensions. 

The gauge sector of QED$_3$ is found to have a mass and a vacuum
energy that are independent of $\theta$ for weak coupling in the limit
of large UV cut-off.  The independence from $\theta$ of the vacuum
energy of the
gauge sector is in agreement with \cite{verg}.  The
non-perturbative dynamical mass generation for the photon, the vacuum
energy density and the expectation value of the Wilson loop are all in
agreement with \cite{var}.  In both \cite{verg} and \cite{var} QED$_3$
was considered without a scalar field.

Further, we find that the vacuum energy of the scalar field is
dependent upon $\theta$, but that this dependence is suppressed.  It
is in the BPS limit of zero scalar
potential and in the limit of a very massive scalar field (large
scalar potential) that the $\theta$ dependence is most greatly
suppressed.  The $\theta$ dependence is in terms of
$O(z^{\frac{3}{2}})$ in the BPS limit but, in the limit of a very
massive scalar field the scalar sector, and hence the total vacuum energy,
becomes independent of $\theta$ in direct agreement with \cite{verg}. 

A non-perturbative dynamical mass proportional to $\theta$ is
generated for the scalar field which does not disappear in the limit
of zero scalar potential.

It is clear from the calculation of the string tension that the
expectation value of the Wilson loop obeys the area law and leads to
confinement.  Its dependence upon $\theta$ is greatly suppressed for
weak coupling in the limit of large UV cut-off.

An extension of Polyakov's work on confining strings,
\cite{Polyakov2}, has shown that our proposed Lagrangian for low
energy QED$_3$, with a scalar field and a $\theta$ term, is equivalent
to a non-standard string theory.  This string theory is of the same
form as that found by Polyakov to be equivalent to purely gauge field
compact QED$_3$ with a $\theta^2$, but sub-leading in UV momentum
cut-off, modification of the photon mass and a $\theta$ dependent
shift of the shape of the minimal surface.  The modification of the
photon mass is in direct
agreement with our variational calculation.

\newsection{Acknowledgements}
We wish to acknowledge illuminating conversations with A. Kovner,
most particularly about the action of the vortex creation operator.
W. E. B. wishes to thank P.P.A.R.C. for a research studentship.

\subsection*{Appendix}
\appendix
\newsection

In this appendix we shall give details of the evaluation of integrals
over the singular function $\chi_{\nu}$ by considering the example
\begin{equation}
Z_{\nu}^{-1} \int D\chi_{\nu} \partial_i \chi_{\nu} G^{-2} \partial_i
\chi_{\nu} \exp [-\frac{1}{4g^2} \partial_i \chi_{\nu} G^{-1} \partial_i
\chi_{\nu} - \frac{4\theta^2}{\eta^2} \pi^2 \rho K \rho]
\label{examp}
\end{equation}
We use the transformation
\begin{equation}
\partial_i \chi_{\nu} = \epsilon_{ij} \partial_j \psi
\end{equation}
The singularities in $\chi_{\nu}$ are angular functions in two
dimensions and so we can use the standard definition
\begin{equation}
\chi_{\nu} = -\frac{i}{2} \sum_{\alpha , \beta} \left( \log
{(\frac{z-z_{\alpha}}{\overline{z} -\overline{z_{\alpha}}})} - \log
{(\frac{z-z_{\beta}}{\overline{z} -\overline{z_{\beta}}})} \right)
\end{equation}
The form of $\psi$ is therefore
\begin{equation}
\psi = \sum_{\alpha , \beta} \left( \log
(\frac{1}{|z -z_{\alpha}|}) - \log
(\frac{1}{|z -z_{\beta}|}) \right)
\end{equation}
so that $\partial_j^2 \psi = -2\pi \rho$ or, in momentum space, $\psi
= -2\pi \rho k^{-2}$ where $\rho$ is the distribution function of the
vortices or singularities of $\chi_{\nu}$ and is defined as 
\begin{equation}
\rho (x) = \sum_{\alpha , \beta} \delta (x-x_{\alpha}) - \delta
(x-x_{\beta})
\end{equation}
So (\ref{examp}) is now transformed to
\begin{equation}
4 \pi^2 k^{-2} G^{-2} <\rho (x) \rho(y)> = 8 \pi^2 \int
\frac{d^2k}{(2\pi)^2} k^{-2} G^{-2}z
\end{equation}
$<\rho \rho>$ is calculated to $O(z)$ in subsection 4.1.

\newsection

We shall functionally minimise the vacuum energy density with respect
to the scalar and vector propagators to obtain the forms of the masses
and propagators of the the fields.  From (\ref{potential}) we note that 
\begin{eqnarray}
\frac{\delta z}{\delta G(k)} &=& \frac{1}{4g^2}k^{-2}G^{-2}(k)z\\
\nonumber
\frac{\delta z}{\delta K(k)} &=& -\frac{\theta^2}{\eta^2} z
\label{37}
\end{eqnarray}
First we consider the minimization of $\frac{1}{V}<H^{\theta}>$ with
respect to $G(k)$.
\begin{equation}
\frac{\delta <H_A^{\theta}>}{\delta G(k)}=\frac
{1}{4}[k^2-G^{-2}(k)+\frac{4\pi^2}{g^2}(2zk^{-2}G^{-3}(k)
-\frac{\delta z}{\delta G(k)}4\pi^2  \int \frac{d^2
p}{(2\pi)^2}p^{-2}G^{-2}(p))]
\label{38}
\end{equation}
Assuming that at large momenta $G(k) \rightarrow k^{-1}$, the ratio of
the fourth term to the third term in (\ref{38}) is 
\begin{equation}
\frac{\delta z}{\delta G(k)} \frac{4\pi^2 \int \frac{d^2
p}{(2\pi)^2}p^{-2}G^{-2}(p)}{2zk^{-2}G^{-3}(k)} \propto
\frac{\Lambda^2}{g^2 k}
\end{equation}
This is much greater than one, at weak coupling, for any value of $k$
and so we omit the third term from (\ref{38}).
Also using
\begin{equation}
\frac{\delta <H_{\phi^3}^{\theta}>}{\delta G(k)} = \frac{16 \pi^4
\theta^2}{\eta^2} \frac{\delta z}{\delta G(k)} \int \frac{d^2
p}{(2\pi)^2}[1-K^2(p)(p^2+\lambda \eta^2)]
\end{equation}
we obtain the minimization equation
\begin{eqnarray}
0 &=& \frac{1}{4}[k^2-G^{-2}(k)]-\frac{\pi^4}{g^4}k^{-2}G^{-2}(k)z\int
\frac{d^2p}{(2\pi)^2}p^{-2}G^{-2}(p)\\ \nonumber
 &+& \frac{4\theta^2 \pi^4}{\eta^2
g^2}k^{-2}G^{-2}(k)z \int \frac{d^2p}{(2\pi)^2}[1-K^2(p)(p^2+\lambda \eta^2)]
\end{eqnarray}

Now consider the minimization with respect to $K(k)$.
\begin{equation}
\frac{\delta <H_A^\theta>}{\delta K(k)} = -\frac{4\pi^4}{g^2}
\frac{\delta z}{\delta K(k)} \int \frac{d^2
p}{(2\pi)^2}p^{-2}G^{-2}(p)
\label{40}
\end{equation}
\begin{eqnarray}
\frac{\delta <H_{\phi^3}^{\theta}>}{\delta K(k)} &=&
\frac{1}{4}[(k^2+\lambda \eta^2)-K^{-2}(k)-\frac{32\pi^2
\theta^2}{\eta^2}zK(k)(k^2+\lambda \eta^2)\\ \nonumber
&+& \frac{64\pi^4 \theta^2}{\eta^2} \frac{\delta z}{\delta K(k)}
\int \frac{d^2p}{(2\pi)^2}[1-K^2(p)(p^2+\lambda \eta^2)]]
\label{41}
\end{eqnarray}
Assuming that at large momenta $K^2(k) \rightarrow k^{-2} +
k^{-4}a^2$, where $a$ is the constant coefficient of the second term in the
expansion, the ratio of the penultimate to the last term in
$\frac{\delta <H_{\phi}^{\theta}>}{\delta K(k)}$ is
\begin{equation}
\frac{K(k)(k^2+\lambda \eta^2)}{2\frac{\theta^2}{\eta^2}\int
\frac{d^2p}{(2\pi)^2} [1-K^2(p)(p^2+\lambda \eta^2)]} \propto
\frac{k(1+\lambda \eta^2
k^{-2})}{\frac{\theta^2}{\eta^2}(a^2+\lambda \eta^2)\log \Lambda}
\end{equation}
This is much less than one for non-zero $\theta$ in the UV limit for
any value of $\lambda$ and so the penultimate term is ignored.  So we
obtain another simple minimization equation
\begin{equation}
0=\frac{1}{4}[\frac{16 \pi^4 \theta^2}{g^2 \eta^2}z\int
\frac{d^2p}{(2\pi)^2}[p^{-2}G^{-2}(p)-\frac{4\theta^2
g^2}{\eta^2}(1-K^2(p)(p^2+\lambda \eta^2))]+(k^2+\lambda \eta^2)-K^{-2}(k)]
\end{equation}

\renewcommand{\footnotesize}{\small}

\noindent

\bigskip

{\renewcommand{\Large}{\normalsize}

\end{document}